\documentclass[pdftex,twocolumn,epjc3]{svjour3}          

\RequirePackage[T1]{fontenc}

\smartqed  

\RequirePackage{graphicx}
\RequirePackage{mathptmx}      
\RequirePackage{flushend}
\usepackage{bm}
\usepackage{epsfig}
\usepackage{amsmath,amsfonts}
\RequirePackage[numbers,sort&compress]{natbib}
\RequirePackage[colorlinks,citecolor=blue,urlcolor=blue,linkcolor=blue]{hyperref}

\journalname{Eur. Phys. J. C}

\begin{document}

\title{Exact cosmological solutions for MOG\thanksref{t1}}
\author{Mahmood Roshan\thanksref{e1,addr1}}
\thankstext[$\star$]{t1}{MOdified Gravity theory}
\thankstext{e1}{e-mail: mroshan@um.ac.ir}
\institute{Department of Physics, Ferdowsi University of Mashhad, P.O. Box 1436, Mashhad, Iran \label{addr1}}
\date{Received: date / Accepted: date}

\maketitle

\begin{abstract}
We find some new exact cosmological solutions for the covariant scalar-tensor-vector gravity theory, the so-called MOdified Gravity (MOG). The exact solution of the vacuum field equations has been derived.  Also, for non vacuum cases we have found some exact solutions with the aid of the Noether symmetry approach. More specifically, the symmetry vector and also the Noether conserved quantity associated to the point-like Lagrangian of the theory have been found. Also we find the exact form of the generic vector field potential of this theory by considering the behavior of the relevant point-like Lagrangian under the infinitesimal generator of the Noether symmetry. Finally, we discuss the cosmological implications of the solutions.
\end{abstract}

\section{Introduction}
We study a Scalar-Tensor-Vector theory which has been introduced as an alternative to particle dark matter \cite{moffat}. This theory is referred as MOG (MOdified Gravity) in the current literature. Mathematically, MOG is complicated than GR in the sense that it postulates more gravitational filed than GR. In fact, MOG is a Scalar-Tensor-Vector theory of gravity, while GR is a tensor theory. In other words, in MOG in addition to the metric tensor, there are two scalar fields ($\mu(x^{\alpha})$ and $G(x^{\alpha})$) and also a massive Proca vector field $\phi^{\alpha}$. In fact the gravitational constant is variable as is in scalar-tensor theories of gravity. Therefore, the existence of these fields will provide some degrees of freedom which may help to handle the dark matter problem without invoking dark matter particles. It is worth mentioning that adding non-minimal scalar fields to the effective Lagrangian of gravitational filed dates back to Brans-Dicke theory \cite{BD}. In this theory the non-minimal scalar filed was introduced in order to incorporate the Mach's principle into GR. However, there are also other motivations for adding into play the scalar fields. For example, non-minimally coupled terms between scalar fields and geometric quantities, such as the Ricci scalar $R$, have to be added to the gravitational action when quantum corrections are taken into account. Also scalar-tensor theories show up as the dimensionally reduced effective theories of higher
dimensional theories, such as Kaluza-Klein theory. For a comprehensive review of the subject, we refer the reader to \cite{caponew}. 

On the other hand, the existence of the vector fields in the gravitational action has its own long story. In vector-tensor theories, in the form of Einstein-ather theories \cite{ea}, there is a Lorentz-violating vector field. The existence of this vector field can dramatically affect the cosmology. For example, it may leave an imprint on perturbations in the early universe \cite{lim}, and it can even affect the growth rate of structure in the universe. Also the vector fields are a simple and natural candidates to explain some certain anomalies in the Cosmic Microwave Background \cite{uzan 2014}. For another example of theories containing vector fields, one may refer to Tensor-Vector-Scalar theory (TeVeS) \cite{bekenstein}. TeVeS is a relativistic covariant theory for Modified Newtonian Dynamics (MOND) \cite{milgrom}. TeVeS has both types of above mentioned fields, i.e. scalar and vector fields.

As we mentioned before, MOG is also another example of theories where scalar and vector fields participate in the gravitational sector. This theory has been applied to explain the rotation curves of spiral galaxies and the mass discrepancy in the galaxy clusters \cite{rot}-\cite{rot6}. Recently, it has been claimed that this theory is in excellent agreement with the rotation curve data for the Milky Way, while MOND does not fit the data \cite{Moffnew}. In fact, the observed rotation curve data for the Milky Way extends as far as $200kpc$ from the center of the galaxy. Data for such a large distances from the core of the galaxy may provide a critical test of modified theories of gravity \cite{Moffnew}.

There are some papers in the relevant literature which have studied the consequences of MOG at the cosmological scale, for example see \cite{cosmology}-\cite{cosmology5}. However, to the best of our knowledge, the astrophysical aspects of this theory has been investigated more extensively than its cosmological consequences. For example, there is no exact cosmological solution for this theory in the literature. This point motivated us to study the cosmological behavior of this theory by looking for exact solutions for the modified Friedmann equations. To do so, we use the \textit{Noether symmetry approach} \cite{ns}. In fact, as every conservation theorem, the Noether conservation theorem allows to reduce dynamics of the cosmological model and gives insight into conserved quantities. This approach has been extensively used in various cosmological models and modified theories of gravity, for example see \cite{ns2}-\cite{ns26}. In \cite{ns26} the Noether symmetry approach is used in non-minimal scalar-tensor theories and in  higher order theories of gravity. Also a general review of the Noether symmetry approach can be found in this paper. In the current paper, i.e. in the context of MOG, there is also a non-minimal scalar field. We shall see that there are some cosmological exact solutions in MOG which correspond to some specific solutions presented in \cite{ns26}, for example see subsection \ref{secnew}. It is worth mentioning at this point that the\textit{ Hojman's conservation theorem} \cite{hojman} has also been used to study the dynamics of cosmological models and to find exact cosmological solutions for them \cite{me}.

The layout of the paper is the following. In Section (\ref{2}), we briefly
review MOG's field equations and derive the corresponding modified Friedmann equations. In section (\ref{3}), we derive the point-like Lagrangian which is needed for applying the Noether symmetry approach. Section (\ref{4}) is devoted
to the application of the Noether conservation theorem to MOG. In this section we find the explicit form of the generic vector field potential by symmetry issues. In Section (\ref{5}), for the first time in the literature, we find some new exact solutions for MOG and discuss their cosmological implications. Conclusions are drawn in Section (\ref{6}).
 \section{Modified gravity (MOG)}
 \label{2}
We start with the generic action of MOG
\begin{eqnarray}\label{action}
&S=\frac{1}{16\pi}\int \sqrt{-g}d^4x [\frac{\chi^2}{2}(R-2\Lambda)+\frac{1}{2}g^{\mu\nu}\nabla_{\mu}\chi\nabla_{\nu}\chi\\&+ \frac{\chi^2}{4}g^{\mu\nu}\nabla_{\mu}\psi\nabla_{\nu}\psi+\omega_0\left[\frac{1}{4}B_{\mu\nu}B^{\mu\nu}+V_{\phi}\right]]+S_M\nonumber
\end{eqnarray}
 where $R$ is the Ricci scalar, $S_M$ is the matter action, $\Lambda$ is the cosmological constant, $\omega_0$ is a positive coupling constant, $B_{\mu\nu}=\nabla_{\mu}\phi_{\nu}-\nabla_{\nu}\phi_{\mu}$, and the scalar fields $\chi$ and $\psi$ are related to those of the original paper \cite{moffat} as $\chi^2=2/G$ and $\psi=\ln \mu$. Also $V_{\phi}$ is the self-interaction potential for the vector field $\phi^{\mu}$ and can be a function of $\phi_{\mu}\phi^{\mu}$ and $\psi$. Variation of (\ref{action}) with respect to $g^{\mu\nu}$, $\phi_{\alpha}$, $\chi$, $\psi$  leads to the following field equations respectively \cite{us2}
 \begin{eqnarray} \label{mog11}
 G_{\mu\nu}+\Lambda g_{\mu\nu}=\frac{1}{\chi^2}(\nabla_{\mu}\nabla_{\nu}-g_{\mu\nu}\square)\chi^2+\frac{16\pi}{\chi^2} T^{total}_{\mu\nu}
 \end{eqnarray}
 \begin{eqnarray}\label{mog12}
 \nabla_{\mu}B^{\alpha\mu}+ \frac{\partial V_{\phi}}{\partial \phi_{\alpha}}=-\frac{16\pi}{\omega_0} J^{\alpha}
 \end{eqnarray}
 \begin{eqnarray}\label{mog13}
    \square \chi=\chi R+\frac{\chi}{2}g^{\mu\nu}\nabla_{\mu}\psi\nabla_{\nu}\psi
 \end{eqnarray}
 \begin{eqnarray}\label{mog14}
 \square \psi=-\frac{2}{\chi}\nabla_{\gamma}\chi\nabla^{\gamma}\psi+\frac{2\omega_0}{\chi^2}\frac{\partial V_{\phi}}{\partial \psi}
 \end{eqnarray}
 Where $G_{\mu\nu}$ is the Einstein tensor and $J^{\alpha}$ is a "fifth force" matter current defined as
 \begin{eqnarray}\label{current1}
J^{\alpha}=-\frac{1}{\sqrt{-g}}\frac{ \delta S_M}{\delta \phi_{\alpha}}
 \end{eqnarray}
 nonzero $J^{\alpha}$ means that there is a coupling between matter and the Proca vector field $\phi^{\mu}$. Consequently, one can verify that the matter energy-momentum tensor is not conserved and the Einstein's equivalence principle is violated \cite{me2}. Also, the total energy-momentum tensor is defined as
\begin{eqnarray}
T^{total}_{\mu\nu}=T_{\mu\nu}+T^{\phi}_{\mu\nu}+T^{\chi}_{\mu\nu}+T^{\psi}_{\mu\nu}
\end{eqnarray}
where $T_{\mu\nu}$ is the energy-momentum tensor for the ordinary matter, and
\begin{eqnarray}\label{mog7}
&T^{\phi}_{\mu\nu}=-\frac{\omega_0}{16\pi}\left(B_{\mu}^{~\alpha}B_{\nu\alpha}-g_{\mu\nu}\left(\frac{1}{4}B^{\rho\sigma}B_{\rho\sigma}+V_{\phi}\right)+2\frac{\partial V_{\phi}}{\partial g^{\mu\nu}}\right)\nonumber\\&
T^{\chi}_{\mu\nu}=-\frac{1}{16\pi}\left(\nabla_{\mu}\chi\nabla_{\nu}\chi-\frac{1}{2} g_{\mu\nu}\nabla_{\alpha}\chi\nabla^{\alpha}\chi\right)\nonumber \\&\
T^{\psi}_{\mu\nu}=-\frac{\chi^2}{32 \pi}\left(\nabla_{\mu}\psi\nabla_{\nu}\psi-\frac{1}{2} g_{\mu\nu}\nabla_{\alpha}\psi\nabla^{\alpha}\psi\right)\nonumber
\end{eqnarray}
Furthermore, we assume a perfect fluid energy-momentum tensor for the ordinary matter. Also we consider the cosmological behavior of MOG in a flat Friedmann-Robertson-Walker (FRW) universe for which the line element is given by
 \begin{eqnarray}
 ds^2=-dt^2+a(t)^2\left(dx^2+dy^2+dz^2\right)
 \end{eqnarray}
 where $a(t)$ is the cosmic scale factor. Now, by inserting this metric to the field equations (\ref{mog11})-(\ref{mog14}), and also noting the relation between $\chi$, $\psi$ and $G$, $\mu$, we find the modified version of the Friedmann equations. the \textit{0-0} component of (\ref{mog11}) gives
 \begin{eqnarray} \label{eq1}
\frac{\dot{a}^2}{a^2}=\frac{8\pi G}{3}\rho+\frac{\Lambda}{3}+&\\&+\left[\frac{\dot{G}}{G}\frac{\dot{a}}{a}-\frac{1}{12}\frac{\dot{\mu}^2}{\mu^2}-\frac{1}{24}\frac{\dot{G}^2}{G^2}+\frac{8\pi G}{3}\kappa \phi_0 \rho\right]\nonumber
 \end{eqnarray}
Where $\phi_0(t)=-\phi^0(t)$ is the zeroth covariant component of the vector field and $\rho$ is the total energy density of the matter/radiation. Also dot denotes the derivative with respect to time. The terms inside the bracket can be considered as corrections to the corresponding standard Friedmann equation. Using the equation of motion of the test  particles in MOG (the generalized version of the geodesic equation in metric theories), one can find an action for a pressure-less dust. Therefore, it can be shown that $J^0=\kappa\rho$ (see \cite{me2} for more detail). Where $\kappa$ is another coupling constant. Although it seems that there are two independent coupling constants $\omega_0$ and $\kappa$, they almost always appear as the combination $\kappa^2/\omega_0$ through the calculations of the physical quantities. Also it should be mentioned that because of the \textit{cosmological principle}, spatial components of $J^{\alpha}$ and all components of antisymmetric tensor $B_{\mu\nu}$ are zero. Furthermore the \textit{i-i} components of (\ref{mog11}) takes the following form
\begin{eqnarray}\label{eq2}
\frac{\ddot{a}}{a}=-\frac{4\pi G}{3}(\rho+3 p)+\frac{\Lambda}{3}+&\\+[\frac{1}{2}\frac{\dot{G}}{G}\frac{\dot{a}}{a}+\frac{1}{6}\frac{\dot{\mu}^2}{\mu^2}+\frac{1}{2}&\frac{\ddot{G}}{G}-\frac{11}{12}\frac{\dot{G}^2}{G^2}-\frac{4\pi G}{3}\kappa\phi_0\rho-\frac{G\omega_0}{6} V_{\phi}]\nonumber
\end{eqnarray}
where $p$ is the pressure of the cosmic fluid, and we have used equation (\ref{eq1}) for terms including $\dot{a}^2$. Equation (\ref{mog12}), i.e. the field equation of the vector field, takes the following form
\begin{eqnarray}\label{eq3}
\frac{\partial V_{\phi}}{\partial \phi_0}=\frac{16\pi\kappa}{\omega_0}\rho
\end{eqnarray}
Therefore the potential $V_{\phi}$ should satisfy the following general criterion for all values of $\mu(t)$ and $\phi_0(t)$
\begin{eqnarray} \label{criterion}
 \frac{\omega_0}{\kappa}\frac{\partial V_{\phi}}{\partial \phi_0}\geq 0
\end{eqnarray}
 Furthermore, equation (\ref{mog13}) can be written as 
\begin{eqnarray}\label{eq4}
\frac{\ddot{G}}{G}=32\pi G(1+\kappa\phi_0)\rho+12\frac{\ddot{a}}{a}+9\frac{\dot{G}}{G}\frac{\dot{a}}{a}-&\\-2\frac{\dot{\mu}^2}{\mu}+\frac{\dot{G}^2}{G^2}-&2G\omega_0 V_{\phi}\nonumber
\end{eqnarray}
and finally equation (\ref{mog14}) reads
\begin{eqnarray}\label{eq5}
\frac{\ddot{\mu}}{\mu}=\frac{\dot{\mu}^2}{\mu^2}-3\frac{\dot{\mu}}{\mu}\frac{\dot{a}}{a}+\frac{\dot{G}}{G}\frac{\dot{\mu}}{\mu}-G\omega_0\mu\frac{\partial V_{\phi}}{\partial \mu}
\end{eqnarray}
If the generic potential $V_{\phi}$ and the equation of state $p=p(\rho)$ are known, then equations (\ref{eq1}),(\ref{eq2}), (\ref{eq3}), (\ref{eq4}) and (\ref{eq5}) are five non-linear partial differential equations for five unknown functions $a(t)$, $\phi_0(t)$, $\mu(t)$, $G(t)$ and $\rho(t)$. It is interesting to remark that the vector field is not "dynamical" in the sense that its field equation does not contain any time derivative. In other words, the vector field is algebraically related to the energy density $\rho$.
\section{Vacuum exact solution}
Before starting the Noether symmetry approach, we consider the vacuum exact solution ($\rho=0$ and $\Lambda=0$). In this case the modified Friedman equations can be analytically integrated for a wide range of generic potential $V_{\phi}$. Let us rewrite Equations (\ref{eq1}),(\ref{eq2}), (\ref{eq3}), (\ref{eq4}) and (\ref{eq5}), respectively, as
\begin{eqnarray}\label{91}
\frac{1}{12}(\frac{\mu'^2}{\mu^2}+\frac{G'^2}{2G^2})=\frac{G'}{G}-1
\end{eqnarray}
\begin{eqnarray}
\frac{H'}{H}=\frac{G'}{G}-3
\end{eqnarray}
\begin{eqnarray}
\mu''=\frac{\mu'^2}{\mu}
\end{eqnarray}
\begin{eqnarray}\label{92}
G''=\frac{G'^2}{G}
\end{eqnarray}
where $H=\dot{a}/a$ and prime denotes derivative with respect to $\ln a$. Also it should be mentioned that we have assumed that the Proca potential has a general form $V_{\phi}=f_1(\mu)f_2(\phi_{\alpha}\phi^{\alpha})$ where $f_1$ and $f_2$ are arbitrary functions. $f_2$ satisfies $f_2(0)=0$ and $\frac{df_2(0)}{d\phi_0}=0$. Note that in the absence of matter/radiation the vector field freezes out at a finite value. We have assumed models for which the vector field freezes at $\phi_0=0$. This is the case for original potential of MOG, i.e. $V_{\phi}=-\frac{1}{2} \mu^2 \phi^{\alpha}\phi_{\alpha}$.

 The exact solutions to equations (\ref{91})-(\ref{92}) are 
 \begin{eqnarray}
 G(t)=G_0\,a(t)^{\epsilon}
 \end{eqnarray}
  \begin{eqnarray}
\mu(t)=\mu_0\,a(t)^{\frac{\pm\sqrt{24\epsilon -24-\epsilon^2}}{\sqrt{2}}}
 \end{eqnarray}
   \begin{eqnarray}
a(t)=(\epsilon_1\, t+\epsilon_2)^{\frac{1}{3-\epsilon}}
 \end{eqnarray}
 where $G_0$, $\mu_0$, $\epsilon$, $\epsilon_1$ and $\epsilon_2$ are constants of integration. One may consider this solution as a late time solution where the energy density of matter and radiation is zero. If $2<\epsilon<3$ then we have an accelerated expansion, i.e. $\ddot{a}>0$. In this case $G(t)$ is an increasing function of time and $\mu(t)$ can be decreasing or increasing. It is important mentioning that this late time acceleration is not a de Sitter universe where the cosmic scale factor grows as $a(t)\sim e^{\sqrt{\Lambda} t}$. 
\section{Canonical point-like Lagrangian for MOG}
\label{3}
Equations (\ref{eq1}),(\ref{eq2}), (\ref{eq3}), (\ref{eq4}) and (\ref{eq5}) can also be deduced from a canonical point-like Lagrangian $\mathcal{L}(a,\dot{a},\chi,\dot{\chi},\phi_0,\dot{\phi_0},\psi,\dot{\psi})$. For notational simplicity, in this section we use $\psi$ and $\chi$ instead of $\mu$ and $G$. The point-like Lagrangian $\mathcal{L}$ is derived from action (\ref{action}). We denote the configuration space by $\mathcal{Q}\equiv\lbrace a,\chi,\psi,\phi_0\rbrace$. Then $\mathcal{TQ}\equiv\lbrace a,\dot{a},\chi,\dot{\chi},\phi_0,\dot{\phi_0},\psi,\dot{\psi}\rbrace$ is the corresponding tangent space where the Lagrangian $\mathcal{L}$ is defined. On the other hand, the Euler-Lagrange equation of $\mathcal{L}$ is
\begin{eqnarray}\label{el}
&\frac{d}{dt}\frac{\partial \mathcal{L}}{\partial \dot{q}}-\frac{\partial \mathcal{L}}{\partial q}=0,~~~
\frac{d}{dt}\frac{\partial \mathcal{L}}{\partial \dot{\phi_0}}-\frac{\partial \mathcal{L}}{\partial \phi_0}=16\pi a^3 \kappa\rho
\end{eqnarray}
where $q$ can be $a$, $\chi$ or $\psi$. Also the energy equation is 
\begin{eqnarray}\label{energy}
E_{\mathcal{L}}= \dot{a}\frac{\partial \mathcal{L}}{\partial\dot{a}}+\dot{\chi}\frac{\partial \mathcal{L}}{\partial\dot{\chi}}+\dot{\psi}\frac{\partial \mathcal{L}}{\partial\dot{\psi}}+\dot{\phi_0}\frac{\partial \mathcal{L}}{\partial\dot{\phi_0}}-\mathcal{L}=0
\end{eqnarray}
therefore combining equations (\ref{el}) and (\ref{energy}), we have five equations which make a complete set of partial differential equations for five unknown functions. 

In order to find $\mathcal{L}$ one must insert the flat FRW metric into action (\ref{action}) and simplify the result using an integration by parts. Note that the Ricci scalar curvature takes the form $R =6(\dot{a}^2/a^2+\ddot{a}/a)$. As a result, one can easily verify that, in a FRW manifold, the Lagrangian related to the action (\ref{action}) takes the point-like form
\begin{eqnarray}\label{pl}
\mathcal{L}=6\chi\dot{\chi}\dot{a}a^2+3a\dot{a}^2\chi^2+\Lambda a^3\chi^2+\frac{1}{2}a^3\dot{\chi}^2+&\\ \frac{1}{4}a^3\chi^2\dot{\psi}^2-\omega_0 a^3 V_{\phi}+16\pi\rho a^3& \nonumber
\end{eqnarray}
where $\rho=\rho_{m}+ \rho_r$ is the proper energy density. Where $\rho_m$ is the matter energy density, and $\rho_{r}$ is the radiation energy density. It is worth mentioning that, in general, the mater energy-momentum tensor in MOG is not conserved (see equation A31 in \cite{me2}). However in the case of FRW universe we have $\nabla_{\mu}T^{\mu\nu}=0$ and consequently one can write $\rho_m=\rho_{m0} \,a^{-3}$ and $\rho_{r}=\rho_{r0}\,a^{-4}$, where $\rho_{m0}$ and $\rho_{r0}$ are the current values of the corresponding energy densities. The current magnitude of the cosmic scale factor has been normalized to unity. It is interesting that although there is a coupling between matter and the vector field, the energy-momentum tensor of the ordinary matter is conserved. This is an enormous simplification made by the cosmological principle.  

Now by writing the Euler-Lagrange equations (\ref{el}) for the Lagrangian (\ref{pl}), one can straightforwardly derive the equations (\ref{eq2})-(\ref{eq5}). On the other hand, the energy condition (\ref{energy}) will give the equation (\ref{eq1}).

\section{Noether symmetry approach}
\label{4}
Now let us introduce the \textit{lift vector field} $\mathbf{X}$ \cite{marmo}. $\mathbf{X}$ is an infinitesimal
generator of the Noether symmetry in the tangent space $\mathcal{TQ}$, and is defined as
\begin{eqnarray}
X=\alpha\frac{\partial}{\partial a}+\beta\frac{\partial}{\partial \chi}+\gamma\frac{\partial}{\partial \psi}+\xi\frac{\partial}{\partial \phi_0}+\dot{\alpha}\frac{\partial}{\partial \dot{a}}+&\\ +\dot{\beta}\frac{\partial}{\partial \dot{\chi}}+\dot{\gamma}\frac{\partial}{\partial \dot{\psi}}+\dot{\xi}\frac{\partial}{\partial \dot{\phi_0}}&\nonumber
\end{eqnarray}
where the functions $\alpha$, $\beta$, $\gamma$ and $\xi$ depend on the configuration space variables $a$, $\chi$, $\psi$ and $\phi_0$. One can straightforwardly show that
\begin{eqnarray}
\frac{d\Sigma}{dt}=L_{\mathbf{X}}\mathcal{L}+16\pi\xi a^3\kappa \rho
\end{eqnarray}
where $L_{\mathbf{X}}\mathcal{L}$ is the Lie derivative of the Lagrangian along the lift vector $\mathbf{X}$, and $\Sigma$ is defined as
\begin{eqnarray}\label{n5}
\Sigma=\alpha\frac{\partial \mathcal{L}}{\partial\dot{a}}+\beta\frac{\partial \mathcal{L}}{\partial\dot{\chi}}+\gamma\frac{\partial \mathcal{L}}{\partial\dot{\psi}}
\end{eqnarray}
Therefor one may conclude that $\Sigma$ is a conserved quantity if  $L_{\mathbf{X}}\mathcal{L}=-16\pi\xi a^3\kappa \rho$ (this is the Noether's theorem). It is needed to recall that in theories which there is no coupling between matter and the gravitational fields, the right hand side of this criterion is zero. By applying this condition to the Lagrangian, the analytic form of the functions $\alpha$, $\beta$, $\gamma$ and $\xi$ will be specified. In fact, one may simply equate to zero the coefficients of terms like $\dot{\psi}^2$, $\dot{\chi}^2$, $\dot{\psi}\dot{\chi}$ and so on in order to find some differential equations for the lift vector's components. This is a common procedure in the relevant literature of the Noether symmetry approach, for example see \cite{ns}-\cite{ns26}. Applying this procedure to the Lagrangian (\ref{pl}), we find the following differential equations
\begin{eqnarray}\label{n1}
&12\chi\frac{\partial\alpha}{\partial\psi}+12a\frac{\partial\beta}{\partial\psi}+\chi a^2\frac{\partial\gamma}{\partial a}=0\nonumber\\&
\beta a+\frac{a^2}{6}\frac{\partial\beta}{\partial a}+\chi^2\frac{\partial\alpha}{\partial\chi}+a\chi\frac{\partial\beta}{\partial\chi}+a\chi\frac{\partial\alpha}{\partial a}+2\alpha\chi=0\nonumber\\&
\frac{3}{2}\alpha+6\chi\frac{\partial\alpha}{\partial\chi}+a\frac{\partial\beta}{\partial\chi}=0\\&
\chi \alpha+2 a \chi \frac{\partial \alpha}{\partial a}+2 a \beta+2a^2\frac{\partial\beta}{\partial a}=0\nonumber\\&
3\alpha\chi+2a\chi\frac{\partial\gamma}{\partial\psi}+2a\beta=0\nonumber\\&
12\chi\frac{\partial\alpha}{\partial\psi}+2a\frac{\partial\beta}{\partial\psi}+a\chi^2\frac{\partial\gamma}{\partial\chi}=0\nonumber\\&
\frac{\partial \gamma}{\partial\phi_0}=\frac{\partial \alpha}{\partial\phi_0}=\frac{\partial \beta}{\partial\phi_0}=\frac{\partial \xi}{\partial\phi_0}=0\nonumber
\end{eqnarray}
Furthermore, there are some terms in $L_{\mathbf{X}}\mathcal{L}$ which do not contain any time derivatives of configuration space variables. The summation of these terms should be separately zero:
\begin{eqnarray}\label{n2}
\left(3\omega_0 V_{\phi}+16\pi\rho_r\right)\alpha-2\Lambda\chi a\left(\beta+\frac{3}{2}\frac{\chi}{a}\alpha\right)+&\\\omega_0 a \frac{\partial V_{\phi}}{\partial\psi}\gamma=0 &\nonumber
\end{eqnarray}
It is obvious from equations (\ref{n1}) that the function $\xi$ can be an arbitrary function of $a$, $\chi$ and $\psi$. In other words, the existence of the Noether symmetry does not restrict the functional form of $\xi$. Therefore, one can construct infinite number of symmetry generators for this cosmological model. It is also worth mentioning that, there are nine differential equations in (\ref{n1}) which are enough for finding the exact form of $\alpha(a,\chi,\psi)$, $\beta(a,\chi,\psi)$, $\gamma(a,\chi,\psi)$. After finding this functions and inserting them into (\ref{n2}), we shall find the form of the generic potentials $V_{\phi}$ for which the Noether symmetry exists. After some straightforward algebraic calculations, one can verify that the general solutions to equation (\ref{n1}) are
\begin{eqnarray}\label{n3}
\alpha &=&(\alpha_1\psi+\alpha_2)a\\ \beta &=&-\frac{3}{2}\frac{\chi}{a}\alpha \nonumber\\ \gamma &=& 6\alpha_1\ln \left(a\chi^{-3/2}\right)+\alpha_3  \nonumber
\end{eqnarray}
Where $\alpha_1$, $\alpha_2$ and $\alpha_3$ are integration constants. Although the Lagrangian (\ref{pl}) seems complicated, it possesses simple Noether symmetry generators. Furthermore, using the second equation in (\ref{n3}), the term containing the cosmological constant $\Lambda$ vanishes in (\ref{n2}). Therefore, equation (\ref{n2}) can be rewritten as  
\begin{eqnarray}\label{n4}
V_{\phi}+\frac{16\pi\rho_{r}}{3\omega_0}+\frac{6\alpha_1\ln\left(a\chi^{-3/2}\right)+\alpha_3}{3(\alpha_1\psi+\alpha_2)}\frac{\partial V_{\phi}}{\partial \psi}=0
\end{eqnarray}
In order to find the exact form of the potential $V_{\phi}$ it is needed to find the scale factor $a(t)$ and the scalar field $\chi$ with respect to $\psi$ and $\phi_0$. To do so, let us first derive the conserved quantity $\Sigma$ associated to the Noether symmetry generator $\mathbf{X}$. We recall that Noether's theorem states that every symmetry of the Lagrangian implies the existence of a conservation law, and therefore a conserved quantity. Using the Lagrangian (\ref{pl}) and equation (\ref{n5}), we find
\begin{eqnarray}\label{constant}
\Sigma=(9 a^2\chi\dot{\chi}-6\chi^2\dot{a}a)\alpha+a^3\chi^2\dot{\psi}\gamma
\end{eqnarray}
For the sake of simplicity we assume that $\Sigma=0$. Thus one can easily verify that when $\alpha_1\neq 0$, we have
\begin{eqnarray}\label{co1}
\ln(a\chi^{-\frac{3}{2}})=\alpha_4\alpha_2-\frac{\alpha_3}{6\alpha_1}+\alpha_4\alpha_1\psi 
\end{eqnarray}
where $\alpha_4$ is a new integration constant. When $\alpha_1$ is zero (i.e. when $\gamma$ is constant), it is easy to show that
\begin{eqnarray}\label{co2}
\ln(a\chi^{-\frac{3}{2}})=\alpha_4+\frac{\alpha_3}{6\alpha_2}\psi 
\end{eqnarray}
Substituting these results into equation (\ref{n4}), we find
\begin{eqnarray}\label{n7}
V_{\phi}+\frac{16\pi\rho_{r}}{3\omega_0}+A^{-1}\frac{\partial V_{\phi}}{\partial \psi}=0
\end{eqnarray}
where $A$ is a constant and is equal to $\frac{1}{2\alpha_1\alpha_4}$ when $\alpha_1\neq 0$ and $A=(3\alpha_2)/\alpha_3$ when $\alpha_1$ is zero. It is important to note that, in principle, one may find the cosmic scale factor as a function of $\frac{\partial V_{\phi}}{\partial\phi_0}$ from the field equation of the vector field (i.e. equation (\ref{eq3})). Therefore, equation (\ref{n7}) is a partial differential equation for potential $V_{\phi}(\phi_0,\psi)$. This equation will determine the general form of $V_{\phi}$ for which the Noether symmetry exists. 
\subsection{The case $\rho_r=0$ and $\rho_m$, $\Lambda \neq 0$}
In this case the energy budget consists of the cosmological constant $\Lambda$, non-relativistic matter, the scalar fields $G$ and $\mu$, and the vector field $\phi^{\mu}$. Equation (\ref{n7}) takes a simple form and can be easily integrated. The result is
\begin{eqnarray}\label{po1}
V_{\phi}(\mu,\phi_0)=f(\phi_0)\mu^{-A}
\end{eqnarray}
where $f(\phi_0)$ is an arbitrary function of $\phi_0$, and we have used $\psi(t)=\ln(\mu(t))$. This potential covers a wide range of potentials. For example the original potential of MOG, i.e. $V_{\phi}=-\frac{1}{2}\mu^2 \phi_{\nu}\phi^{\nu}$ \cite{moffat}, lies in this subclass. Another, straightforward example is
\begin{eqnarray}
V_{\phi}(\mu,\phi^{\alpha})=V_0 \mu^n\left(-\phi_{\nu}\phi^{\nu}\right)^{m/2}
\end{eqnarray}
which in an isotropic and homogeneous space-time takes the form $V_{\phi}=V_0 \mu^n\phi_0^m$. Also, it is obvious that if $\alpha_2$=0 (or equivalently $A=0$) then $V_{\phi}$ is a function of the vector field and does not depend on the scalar field $\mu$. More importantly, every function of $\phi_{\mu}\phi^{\mu}$ is a solution. 
\subsection{The case $\rho_m=0$ and $\rho_r$, $\Lambda \neq 0$}
\label{new}
In this case, we rewrite equation (\ref{n7}) as follows
\begin{eqnarray}\label{n8}
V_{\phi}+\frac{1}{3\kappa}\frac{\partial V_{\phi}}{\partial \phi_0}+A^{-1}\frac{\partial V_{\phi}}{\partial \psi}=0
\end{eqnarray}
and the general form of the solution is
\begin{eqnarray}\label{po2}
V_{\phi}=f(y)\mu^{-A}
\end{eqnarray}
in which $y=\exp(\phi_0)\mu^{-A/3\kappa}$, and $f$ is an arbitrary function of $y$. Simple examples of this class of potentials, are $V_{\phi}=V_0 \mu^{\lambda_1}$ and $V_{\phi}=V_0 \exp(\lambda_1 \sqrt{-\phi_{\alpha}\phi^{\alpha}})\mu^{\lambda_2}$. 
\section{Cosmological exact solutions for MOG}
\label{5}
Now we present some new exact cosmological solutions to MOG's modified Friedmann equations. Also we restrict ourselves to solutions with zero radiation density. We recall that by setting the Noether constant of motion $\Sigma$ to zero, we found  equations (\ref{co1}) and (\ref{co2}). Let us rewrite these equations as
\begin{eqnarray}\label{co3}
a(t)=c G(t)^{-3/4}\mu(t)^{1/2A}
\end{eqnarray}
where $c$ is a constant. This is a key equation to derive exact solutions. Using this equation we find some solutions for specific potentials introduced in the previous sections as potentials for which the Noether symmetry exists.
\subsection{ Model $V_{\phi}=\lambda_0 \mu(t)^{2q}$, $\rho_m=\rho_r=\Lambda=0$}
Where $\lambda_0$ is a constant. We have used a new parameter $q$ instead of $A$ as $A=-2q$. This model corresponds to a scalar-tensor theory with two scalar fields $\mu(t)$ and $G(t)$. In fact, since the matter and radiation are absent in the model, there is no source for producing the vector field. With the aid of equation (\ref{co3}) and assuming a power-law type solutions for the scalar fields as $\mu(t)=\mu_0 t^n$ and $G(t)=G_0 t^m$, we find the following exact solution
\begin{eqnarray}\label{so1}
&a(t)= a0\,{t}^{ \frac{1}{4-20\,{q}^{2}}}\nonumber\\&G(t)=G0\,{t}^{ \frac{2}{5\,{q}^{2}-1}}\\&
\mu(t)=\mu_0\,{t}^{{\frac {5q}{1-5\,{q}^{2}}}}\nonumber
\end{eqnarray}
where $G_0$ is related to other constants as
\begin{eqnarray}
G_0=-\frac{5}{8}\,{\frac {20\,{q}^{2}+7}{\omega_0\,\lambda_0\,{ \mu_0}^{2q}
 \left( 25\,{q}^{4}-10\,{q}^{2}+1 \right) }}
\end{eqnarray}
One can easily show that this solution corresponds to an accelerated expansion (i.e. $\ddot{a}>0$ and $\dot{a}>0$) if $\sqrt{3/20}<q<1/\sqrt{5}$ or $-1/\sqrt{5}<q<-\sqrt{3/20}$. In this case, the magnitude of the scalar $G(t)$ decreases with time and for $\mu(t)$ increases. 

This solution has an interesting feature: the model can be contracting (i.e. $\dot{a}<0$) while $\ddot{a}>0$. This situation happens when $q>1/\sqrt{5}$ or $q<-1/\sqrt{5}$. In this case $\mu(t)$ is a decreasing function of time and $G(t)$ is increasing. In fact the existence of the scalar fields $\mu(t)$ and $G(t)$ can lead to repulsive gravitational force in the early universe. 
\subsection{ Model $V_{\phi}=-\frac{1}{2}\mu^2 \phi_{\alpha}\phi^{\alpha}$, $\rho_r=\Lambda=0$ and $\rho_m\neq 0$}
As we have already mentioned, this potential is the original potential postulated in MOG \cite{moffat}. For this model we use equation (\ref{co3}) and also assume power-law type solution for $G(t)$ and $\mu(t)$. One can easily verify the following exact solution
\begin{eqnarray}
&a(t)=\sqrt [3]{{\frac {544}{9}}\,{\frac {\pi \,\alpha_0 G_N{ \rho_{m0}}}{{{
 \mu_0}}^{2}{}}}}{t}^{{\frac {12}{23}}}\nonumber\\&
G(t)={\frac {2312}{207}}\,\frac{\alpha_0 G_N}{{{ \mu_0}}^{2}}{
t}^{-{\frac {10}{23}}}\\&
\mu(t)={ \mu_0}{t}^{-{\frac {18}{23}}}\nonumber\\&
\kappa \phi_0(t)={\frac {9}{34}}\nonumber\\&\rho_m(t)={\frac {9}{544}}\,\frac{\mu_0^{2}}{\pi\,\alpha_0 G_N}
{t}^{-{\frac {36}{23}}}\nonumber
\end{eqnarray}
where $\alpha_0=\kappa^2/\omega_0 G_N$ is one of the free parameters of MOG, and $G_N$ is the gravitational constant. Its observational values is $\alpha_0=8.89\pm0.34$ \cite{rot4}. One can easily verify that the condition (\ref{criterion}) is satisfied. It is obvious that the "velocity" of the vector field is zero and field's value remains constant in this matter dominated phase. Also this solution corresponds a decelerating universe. If we compare this solution with the matter dominated universe in GR (flat FRW space time with $\Lambda=0$) where $a(t)\sim t^{2/3}$, then we see that the scale factor increases slower than the standard Friedman model. Thus the matter density also decreases with slower rate. 
\subsection{ Model $V_{\phi}=V_0\,\mu\, \phi_{\alpha}\phi^{\alpha}$, $\rho_r=\Lambda=0$ and $\rho_m\neq 0$}
Where $V_0$ is a constant. With the aid of equation (\ref{co3}), one may easily verify the following solution
\begin{eqnarray}
&a(t)={ a_0}\,{t}^{-\frac{1}{3}\,m}\nonumber\\&G(t)={ G_0}\,{t}^{-2-m}\nonumber\\&
\mu(t)={\frac {{\kappa}^{2} \left( 3\,\pi \,m+10\,\pi  \right) ^{2}}{{ 4 G_0}\,{\pi }^{2}{ V_0}\,m{ \omega_0}}}{t}^{
m}\\&
\kappa\phi_0(t)=-\,{\frac {2\,\pi \,m}{ \left( 3\,\pi \,m+10\,\pi \right)}}\nonumber\\&\rho_m(t)={\frac {3\,\pi \,m+10\,\pi }{{16\, G_0}\,{\pi }^{2} }} {t}^{m}\nonumber
\end{eqnarray}
where $m$ can has two values: $m=-18/7$ and $m=-3$, and $G_0$ is related to the other constants as
\begin{eqnarray}
G_0={\frac { \left( 3\,\pi \,m+10\,\pi  \right) {{a_0}}^{3}}{{16\,\rho_{m0}}\,{\pi }^{2}}}>0
\end{eqnarray}
it is worth mentioning that the condition (\ref{criterion}) holds if $G_0>0$. As we see the vector field is again constant. Also, the existence of two different values for $m$ shows that our power-law type solutions are not the complete and unique solutions to the cosmological field equations. For $m=-3$ the cosmic scale factor increases linearly with time and so without deceleration, i.e. $\ddot{a}=0$. therefore scale factor increases faster than the standard matter dominated Friedmann model. Also we recall that in GR, for a matter dominated universe with flat spatial curvature and zero cosmological constant, $\ddot{a}$ can not vanishes. 

For $m=-18/7$ the scale factor increases as $a(t)\sim t^{6/7}$. therefore the scale factor increases faster than the standard case. However, unlike the case $m=-3$, we have a negative acceleration ($\ddot{a}<0$). As a final remark for this solution, we mention that the repulsive nature of the theory is again evident. In other words, the combination of the scalar fields and the vector field MOG can yield a repulsive gravitational force at the cosmological scale.
\subsection{ Model $V_{\phi}=V_0\sqrt[m]{-\phi_{\alpha}\phi^{\alpha}}\mu^n $, $\rho_r=\Lambda=0$ and $\rho_m\neq 0$}\label{secnew}
Where $m$ and $n$ are real constant, and $m\neq 0,1,2$. Therefore this model is different from those presented in the previous subsections. Again, by using equation (\ref{co3}) which was obtained from the conserved quantity of the Noether symmetry, we find the following solution
\begin{eqnarray}\label{es4}
&a(t)=a_0\,t^2\nonumber\\&G(t)=G_0\,t^4\nonumber\\&\mu(t)=\mu_0\,t^K\\&
\kappa \phi_0(t)=\kappa\Phi_0=\frac{m}{(2-m)}\nonumber\\&\rho_{m}(t)=\frac{5\,(2-m)}{8\pi G_0}\frac{1}{t^6}\nonumber
\end{eqnarray}
where $K=\pm 2\sqrt{15}$ corresponding to $n=\mp \sqrt{15}/5$ and the relations between constants are
\begin{eqnarray}
V_0=\frac{10\,\mu_0^{K/10}\,\Phi_0^{-m}}{\omega_0\, G_0},~G_0=\frac{5\,(2-m)\,a_0^3}{8\,\pi\,\rho_{m0}}
\end{eqnarray}
The criterion (\ref{criterion}) leads to $(2-m)/G_0>0$. If we assume that $G_0>0$ (note that $G(t)$ is related to the gravitational constant), then everything is well defined provided that $m<2$. It is interesting that this solution represent a constant acceleration for the cosmic expansion. We mention again that such a behavior is not achievable in the corresponding standard Friedmann model. As it is clear from equations (\ref{es4}), the scalar field $G(t)$ is an increasing function of time, and the scalar field $\mu(t)$ depending on the value of parameter $K$ can be an increasing function of time or a decreasing one. However, the rate of expansion is independent of the behavior of $\mu(t)$. 

It is worth mentioning that $G$ is a non-minimal scalar field. As we have already mentioned, non-minimal scalar-tensor theories have been already investigated using the Noether symmetry approach, for example see \cite{ns26}. Although there is an extra vector field  in MOG which is coupled to matter distribution, there are some similarities between the exact solutions of these theories. For example the exact solution presented in this subsection is reminiscent of the exact solution presented in equation (6.42) in \cite{ns26}.
\subsection{ Model $V_{\phi}=-\frac{1}{2}\mu^2\,\phi^{\alpha}\phi_{\alpha}+\frac{\mu^2}{\kappa}\sqrt{-\phi^{\alpha}\phi_{\alpha}}+\frac{\mu^2}{2\kappa} $, $\rho_r=\Lambda=0$ and $\rho_m\neq 0$}
In this case, we found the solution exact solution:
\begin{eqnarray}
&a(t)=a_0\,{t}^{-{
\frac {11}{30}}\,m+\frac{1}{3}}\nonumber\\&
G(t)=G_0\,t^{-\frac{m}{5}}\nonumber\\&\mu(t)=\mu_0\,t^{m}\\&
\kappa \phi_0(t)=-1-{\frac {3\,m \sqrt{\alpha_0 G_N}}{\sqrt {10\,{
G_0}\,{{\mu_0}}^{2}}}}\,{t}^{-{\frac {9}{10}}\,m-1}\nonumber\\&
\rho_m(t)=\frac{\rho_{m0}}{a_0^3}\,t^{\frac{11}{10}\,m-1}\nonumber
\end{eqnarray}
where $m=5/4$ or $-20/11$ and 
\begin{eqnarray}
a_0=\frac{2}{3}\sqrt{2}\,3^{2/3}\,5^{1/6}\,\sqrt [6]{\,{\frac {{G_0}\,\alpha_0 G_N{\pi }^{2}{{\rho_{m0}}}^{2}}{{m}^{2}{\mu_0}}}}
\end{eqnarray}
The condition (\ref{criterion}) is satisfied if $\mu_0>0$ and $G_0>0$. For $m=5/4$ the scale factor decreases as $a(t)\sim t^{-9/8}$, and so $\dot{a}<0$ and $\ddot{a}>0$. However, for $m=-20/11$ the scale factor grows as $a(t)\sim t$.
\section{Conclusion}
\label{6}
In this paper we have studied the cosmology of a specific scalar-vector-tensor theory of gravity known as MOG using a well-known approach. We have found the Noether symmetry generators and also the conserved quantity associated to the symmetry. More specifically, this approach helped us to find the explicit form of the vector field potential $V_{\phi}$ (see equations (\ref{po1}) and (\ref{po2})). We showed that the original potential postulated in this theory, is one of those potentials for which the Noether symmetry exists. Also, using the conserved quantity associated to the Noether symmetry, we have found some exact solutions for the modified Friedmann equations of MOG. We have found five set of exact solutions and except in one of them, we have not neglected the ordinary matter contribution. More specifically, our solutions correspond to matter dominated universes. As we mentioned before, power-law type exact solutions are not necessarily unique. However, these solutions are very often used in cosmological theories in order to insight into the physical content of them. In the context of MOG, there are some interesting features in the exact solutions which may help to understand the cosmological behavior of this theory. For example, the existence of the scalar and vector fields in MOG can lead to contracting universe with positive acceleration. In other words, these fields effect can behave like repulsive gravitational force at the cosmological scales. Also, there exist some solutions which the "gravitational constant" (the scalar field $G(t)$) increases with time as well as solutions where $G(t)$ is decreasing. 
\begin{acknowledgements}
This work is supported by Ferdowsi University of
Mashhad under Grant No. 2/32649 (21/11/1393). The author thanks Sarah Jamali for pointing out an error in a previous version of this paper.
\end{acknowledgements}

\end{document}